# Epitaxial graphene on SiC(0001): More than just honeycombs

Y. Qi, S. H. Rhim, G. F. Sun, M. Weinert, and L. Li*

Department of Physics and Laboratory for Surface Studies

University of Wisconsin, Milwaukee, WI 53211

**Abstract:**

Using scanning tunneling microscopy with Fe-coated W tips and first-principles calculations, we show that the interface of epitaxial graphene/SiC(0001) is a warped graphene layer with the periodic inclusion of hexagon-pentagon-heptagon ($H_{5,6,7}$) defects that break the honeycomb symmetry, thereby inducing a gap and states below $E_F$ near the K point. Although the next graphene layer assumes the perfect honeycomb lattice, its interaction with the warped layer modifies the dispersion at the Dirac point. These results explain recent angle-resolved photoemission and carbon core-level shift data, and solve the long-standing problem of the interfacial structure of epitaxial graphene on SiC(0001).

*lianli@uwm.edu

Graphene, a one-atom-thick planar sheet of $sp^2$-bonded carbon atoms densely packed in a honeycomb lattice, is the building block of many carbon allotropes. The strong in-plane σ bonds form the backbone of the honeycomb, while the half-filled π bonds exhibit linear band dispersion near the K points [1]. This unique construct leads to graphene's novel physical and electronic properties such as room temperature quantum Hall effects, Klein tunneling, and high carrier mobility [2-4]. Recently, uniform wafer-sized graphene has been grown epitaxially on hexagonal SiC [5] and transition metal substrates [6,7], an important technological step towards the development of graphene electronics [8,9].

In the case of epitaxial graphene on SiC, profound interfacial effects have been reported. On the (000-1) C-face, the formation of a "twisted" interface leads to the decoupling between different layers of multilayer graphene, each behaving as a single layer with a carrier mobility of 250,000 $cm^2$/Vs [10,11], comparable to that of exfoliated graphene [4]. On the (0001) Si-face, however, the picture remains controversial [12]. Structurally, graphitization has been known since 1975 to start with a $(6\sqrt{3}x6\sqrt{3})$ structure [13], which remains at the interface during subsequent layer growth. While earlier studies suggested that the structure consisted of graphene layers weakly bonded to either the SiC(0001) (1x1) surface [13,14] or Si-rich interface layers [15-17], recent work indicates a carbon layer covalently bonded to the SiC(0001) [18-23]. Another hotly debated issue is the origin of the gap near the K points observed by angle-resolved photoemission spectroscopy (ARPES) – a property crucial for its use in electronic devices [24-26] – but found only for epitaxial graphene on the Si-face, and absent for the C-face and exfoliated graphene.

In this Letter, we show that these unique properties of epitaxial graphene on Si-face arise from a warped interfacial graphene layer, resulting from the periodic inclusion of hexagon-pentagon-heptagon ($H_{5,6,7}$) defects in the honeycomb to relieve the mismatch with the SiC substrate. The $H_{5,6,7}$ defects break the symmetry of the honeycomb, thereby inducing a gap: the calculated band structure of the proposed model along $\Gamma$-K is semiconducting with two localized states near K points below $E_F$, correctly reproducing the published photoemission and C 1s core-level spectra [22,23]. We further show that the next graphene layer assumes the defect-free honeycomb lattice, though its interaction with the warped layer leads to deviations from the linear dispersion at the Dirac point, shedding light onto the origin of the observed anomalies in angle-resolved photoemission spectra [25-26].

Experiments were carried out on epitaxial graphene grown on N-doped 6H-SiC(0001), which was first etched in a $H_2$/Ar atmosphere at 1500 $^o$C. Then the SiC substrate was annealed at ~950 $^o$C for 15 min in a Si flux to produce a (3x3) reconstructed surface, and finally heated to ~1300 $^o$C to grow graphene in ultrahigh vacuum (UHV) [27]. Scanning tunneling microscopy (STM) images were taken using W and Fe-coated W tips, where the latter are made by coating with Fe at room temperature and followed by annealing at 500-700 $^o$C in UHV.

First-principles calculations, using the Full-potential Linearized Augmented Plane Wave (FLAPW) method as implemented in *flair* [28], model the substrate using a 3x3 6H-SiC(0001) 6-bilayer supercell, with a vacuum region of ~20-25 Å and a basis cutoff of ~195 eV . Because of the usual density functional theory underestimation of the gap (1.6 vs, 3.0 eV for 6H-SiC), comparisons of calculated density of states (DOS) and STM images are limited to biases within $E_F\pm0.5$ eV.

Figure 1 presents an STM image of the ($6\sqrt{3}$x$6\sqrt{3}$) graphene/SiC(0001) interface, and close-ups of the two main features observed: an up-pointing trimer marked by a triangle of sides ~ 3.5 $\pm$ 0.2 Å, and a rosette marked by a hexagon of sides~3.2 $\pm$ 0.2 Å. While the appearance of the trimer does not depend on bias, the contrast of the rosette is slightly less in empty-state images. The ratio of the trimer to rosette varies from sample to sample depending on growth conditions, but the former is always more populated. The image shown in Fig. 1(a) was taken on a sample at an early stage in the growth (Si-rich √3 reconstruction still can be seen on part of the surface), and shows a trimer-to-rosette ratio of ~2:1. Higher ratios (e.g., 4:1) are observed on samples at later stages of growth, which also exhibit a greater number of defects. A close examination of the images and line profiles such as that along the dashed line in Fig. 1 indicates: 1) neighboring features do not always fall on the same line; and 2) the spacing between them is not uniform, with an average value of ~19 Å, i.e., about 6 times the (1x1) lattice spacing of SiC(0001). Furthermore, the center of the rosettes consists of a downward pointing trimer, which can be better seen in the 3D image presented in Fig. 2(a).

STM images (not shown) using W tips at larger bias (e.g., $E_F \pm 1.5$ eV) are similar to those reported previously for the (6√3x6√3) reconstruction [12, 29-31], although imaging at energies closer to $E_F$ (e.g., within ±0.1 eV) is challenging, also consistent with earlier studies [29-31]. With Fe-coated W tips, imaging at these lower energies can be routinely achieved. Our modeling of the Fe/W tip indicates that the Fe minority spin channel has an especially sharp peak at 0.5 eV below $E_F$, which facilitates tunneling between the tip and graphene-specific states not accessible using conventional W tips. These images clearly reveal new details of the (6√3x6√3) structure that cannot be explained by existing models. For example, simulated STM images based on covalently bonded defect-free graphene to SiC(0001) (1x1) show features of either three-fold or two-fold symmetry only [20,21]. In addition, most of these calculations indicate that (6√3x6√3) is metallic [18-20], contrary to a semiconducting surface around K as evident in ARPES [22,23]. A highly interacting graphene layer suggested earlier can indeed yield a semiconducting gap [17], but it also requires complexes of Si tetramers and adatoms bonding to the SiC at the graphene/SiC interface. However, ARPES results show that states associated with Si atoms are not present at this stage of graphene growth [22].

Here we propose a new interface model that accounts for STM observations, as well as ARPES and carbon core level data [22-25]. Because a (13x13) graphene lattice is nearly commensurate to the $(6\sqrt{3}x6\sqrt{3})$ SiC(0001) [12], the growth of a graphene layer will result in two types of high symmetry positions: either a carbon atom or a graphene hexagon centered above a Si. Our calculations indicate that C atoms located directly above Si are pulled towards the SiC surface significantly such that the Si-C bond is shortened to 2.0 Å from a nominal interplanar separation of ~2.3 Å, consistent with other calculations [21].

To better accommodate this bond distortion and retain the three-fold coordination for each carbon atom, pentagons and heptagons – which cause positive and negative curvatures [32-34], respectively – can be inserted into the honeycomb lattice. Arranged such that three pairs of alternating pentagons and heptagons surround a rotated hexagon ($H_{5,6,7}$) (Fig. 2(b)), the inclusion of these defects significantly reduces the distortions of the C-C bonds, and preserves the long-range translational and rotational integrity of the graphene honeycomb. Placements of the $H_{5,6,7}$ defects at the two high symmetry positions on SiC lead to two variants: At the "top" site (Fig. 2(c)), three Si atoms sit directly below the corners of the central hexagon of the $H_{5,6,7}$, with this hexagon centered above a C of the SiC substrate. Similarly at the "hollow" sites (Fig. 2(d)), the central hexagon is centered over a Si and three Si atoms are now bonded to C atoms at the boundary between the $H_{5,6,7}$ defect and the honeycomb lattice. Overall, this transformation decreases interfacial Si-C bonds from 4 (6) to 3 at the hollow (top) sites, further reducing the mismatch with the SiC substrate. The result is a warped graphene layer covalently bonded to SiC(0001) (1x1), whose formation is favored by ~0.1 eV/C compared to a (relaxed) honeycomb structure, with the top site more stable than the hollow by 0,03 eV/C. (The calculated adhesion energy of the honeycomb layer, relative to isolated graphene, is slightly non-binding by ~0.02 eV/atom.) These results are in contrast to unsupported graphene where the $H_{5,6,7}$ defect formation energy is ~5.1 eV; the stability of the warped layer on SiC(0001) is due to its ability to accommodate the strain induced by the Si-C interactions.

Calculated local DOS isosurfaces (corresponding to the STM constant current mode) for the two $H_{5,6,7}$ variants on SiC are shown in Figs. 2 (c) and (d). For the top site (Fig. 2(c)), the six-fold symmetry of the center hexagon is broken by the formation of three Si-C bonds, leading to maxima at the adjacent three alternating C atoms, appearing as a trimer of ~3 Å. For the hollow site variant, depressions are seen within the three heptagons as well as at the three C atoms at the tips of the three pentagons, resembling the six-fold depressions (marked by the hexagons in Fig. 1(b) and (c)) seen by STM. Note that because the six depression sites are inequivalent, they do not coincide perfectly with the hexagon of ~3 Å, consistent with STM observations. The central three C atoms above the $T_4$ sites are slightly brighter, appearing as a downward trimer. Overall the main features seen in the STM images are well reproduced in the calculations. (That the slight center depressions seen in the calculated images are not observed in the STM images may be attributed to (*i*) the calculations do not explicitly include the structure of the tip and (*ii*) the STM images are taken at room temperature.)

In our proposed model of the $(6\sqrt{3}x6\sqrt{3})$ structure, the electronic structure is significantly altered. Thus, experimental ARPES results provide a stringent test of our (and other) proposed models [18-21]. Although the system does not have the (1x1) graphene periodicity, the bands can still be "unfolded" into the (larger) graphene Brillouin zone (BZ) by projecting the (supercell) wave functions onto the corresponding $k$ of the (1x1) graphene cell. The calculated $k$-projected surface bands for the two $H_{5,6,7}$ variants on SiC(0001), combined and individually, are given in Figs. 3(a)-(c). Both variants show definite gaps, i.e., are semiconducting along the $\Gamma$-K direction. Compared to defect-free graphene, our model also shows well-developed graphene-like $\sigma$ bands shifted to greater binding energy, but with significant changes in the $\pi$-band region: (1) increased (diffuse) weight around $\Gamma$ and (2) upward dispersing bands that do not reach the Fermi level but are ~3 eV below at K. In both variants, there are defect-induced states (marked by arrows in Fig. 3) with energies at K, $\varepsilon(K)$, of about -0.9 and -1.8 eV for the hollow and top configurations, respectively. The dispersions of these states (whose intensities decrease significantly away from K) have a tight-binding behavior appropriate for a single-band of localized orbitals, $\varepsilon(K) = \varepsilon_H - A$: the period of the oscillations seen in Figs. 3(b) and (c) reflect the 3x3 cell used in the calculations, while the amplitude $A$~0.3 eV about the "on-site" energies $\varepsilon_H$ (indicated by the horizontal lines) are a measure of the interactions among the $H_{5,6,7}$ defects. From tight-binding scaling arguments, the use of larger $6\sqrt{3}x6\sqrt{3}$ (or 6x6) cells would result in a decrease in $A$ (and interactions) by at least an order of magnitude, i.e., our model predicts (almost dispersionless) states around K with energies $\varepsilon_H \approx$ -0.6 and -1.5 eV, in excellent agreement with the experimental values of 0.5 and 1.6 eV [22,23]. The calculated bands along $\Gamma$-M similarly are in good agreement with the experimental ARPES spectra.

The apparent inconsistency, both experimentally and theoretically, between a gap along $\Gamma$-K and STM imaging at low bias can be understood by noting that STM probes the electronic states in the outer tails of the electronic distribution, whereas ARPES is sensitive to the overall wave function. In Figs. 3(d) and (e), the $k$-projected bands along $\Gamma$-M-K are shown for the two variants, but now weighted by the contributions in the vacuum region probed by STM. In both cases, the gap at K is still seen, but states elsewhere in the (1x1) zone cross $E_F$ (such as those circled in 3(d) and (e)), and are responsible for the contrast seen in STM images at low biases.

In addition to comparisons to valence band photoemission, we present calculated initial state C 1s core level shifts in Fig. 3(f), separated into contributions from the SiC substrate, the carbon atoms of the $H_{5,6,7}$ defect (~-1.1 eV), and the rest of the C atoms (-1.8 eV), consistent with experimental results [22,23]. Over all, the qualitative and quantitative agreement with the available experimental measurements for these defining properties of the (6√3x6√3) layer – the gap at the Fermi level, the overall dispersion, the presence of the two localized states near K, and the carbon core level shifts – provides significant support for our model of the graphene/SiC(0001) interface layer.

The next question is whether the warped interface layer remains during the subsequent growth of graphene multilayers, as shown in Fig. 4(a). Both the honeycomb and the interfacial structures (e.g., the trimers) can be clearly seen on the large terrace, and only a triangular lattice with a 2.5 Å spacing is observed on the smaller terrace in the lower right hand corner. This confirms that the warped graphene layer remains at the interface, and is still accessible by electron tunneling up to the first layer [12,17]. Calculations indicate that the first layer graphene is quite flat, and resumes the perfect honeycomb lattice spacing of 2.5 Å and an interlayer spacing of 3.2 Å relative to the warped interface layer, which still shows significant local buckling. The $k$-projected band structure for the interface + 1[st] layer graphene (Fig. 4(b)) shows almost perfect graphene bands, with the Dirac point below $E_F$. The bottom of the σ (π)-band at Γ is shifted upwards by about 1.3 (3.5) eV compared to the interface (Fig. 3), qualitatively consistent with the ARPES data [22,23].

The bands near the Dirac point are shown in more detail in Figs. 4(c)-(e). The downward shift of $E_D$ = -0.4 eV indicates that the layer is n-doped, consistent with experiment [25,26]. From inspection of the calculated eigenvalues and wave functions, the splitting of the Dirac states is only 33 meV. However, because of the interactions with the warped interface graphene layer, there are deviations from the linear dispersion of defect-free graphene, leading to a parabolic dispersion above the gap, and an apparent gap of ~0.25 eV (marked by the arrows in Fig. 4(c)), closely matching the 0.26 eV gap reported in earlier ARPES studies [25]. Comparison of Figs. 4(d) and (e) reveals subtle, but distinct, differences in the dispersions, especially above the Dirac point. These differences are a direct consequence of the different interactions of the states in the graphene layer with the states of the two $H_{5,6,7}$ variants.

Closely related is the misalignment of the bands above and below $E_D$, illustrated by the dotted lines in Figs. 4(c)-(e): The projections of the $\pi$ states below $E_D$ do not pass through the $\pi^*$ states above $E_D$, an observation previously attributed to electron-phonon or electron-plasmon interactions [26]. (By shifting the lines upward, the dispersion of the $\pi^*$ bands above $E_D$ can be fit, but then the $\pi$ states below $E_D$ are misaligned.) Our results indicate that interactions of the $\pi$ states with the $H_{5,6,7}$ defects contribute significantly to the observed dispersion at the Dirac point.

In summary, the atomic structure of the graphene/SiC(0001) interface is found to be a warped graphene layer with the inclusion of $H_{5,6,7}$ defects in the honeycomb lattice, and the subsequent layer assumes the perfect honeycomb structure. The presence of this interfacial layer, however, modifies its electronic properties. Our results, which provide a consistent explanation of the available experimental data, resolve a long-standing controversy regarding the interfacial structure of epitaxial graphene on SiC(0001), a material that may significantly impact the development of graphene electronics.

**Acknowledgements:** This work was supported by the U.S. DOE, Office of Basic Energy Sciences (DE-FG02-07ER46228).

**Figure captions:**

Fig. 1 (Color online) (a) STM image of epitaxial graphene on 6H-SiC(0001) taken with a Fe-coated W tip at sample bias $V_s$= -0.1 V, tunneling current $I_t$=0.3 nA. Expanded views of marked features taken at (b) -0.1 V and (c) +0.1 V. Image size: 4.5 x4.5 nm$^2$. The arrow marks a commonly seen defect likely associated with Si vacancies in the SiC substrate.

Fig. 2 (Color online) (a) An STM image present in 3D, taken with Fe/W tip at sample bias -0.5 V and $I_t$=0.4 nA, size=4x4 nm$^2$. (b) Proposed model for the $(6\sqrt{3}x6\sqrt{3})$ structure with "top" and "hollow" variants. The $H_{5,6,7}$ defects can be placed into the graphene lattice without causing dislocations, thus naturally allowing for disordered arrangements, consistent with STM observations. Calculated DOS isosurfaces ($10^{-6}$ a.u.$^{-3}$) for occupied states between -0.1 eV and the Fermi level for the (c) top and (d) hollow variants. Carbon atoms are represented by small balls; Si, by larger balls.

Fig. 3 (Color online) Calculated $k$-projected surface bands (convoluted by a decaying exponential to account for the photoelectron escape depth) for graphene with $H_{5,6,7}$ defects on SiC(0001) along Γ-K for (a) equal contributions of both variants, and (b) top and (c) hollow variants separately. Arrows mark the localized states at K and the lines in (b) and (c) indicate $\varepsilon_H$. Vacuum-weighted bands along Γ-M-K for (d) top and (e) hollow variants. Note that the (1x1) graphene BZ corresponds to an extended zone scheme for the supercell. (f) Calculated initial state 1s core level shifts for different carbon atoms.

Fig. 4 (Color online) (a) STM image taken with a W tip at sample bias $V_s$=-0.9 V, tunneling current $I_t$=1.2 nA, image size=15 x 15 nm$^2$. (b) $k$-projected surface bands for the warped interface + 1$^{st}$ layer graphene on SiC(0001). (c) Bands around the Dirac point; (d) top and (e) hollow variants. The arrow at ~ -0.4 eV marks the (split) Dirac point and dotted lines are guides for the linear dispersion.

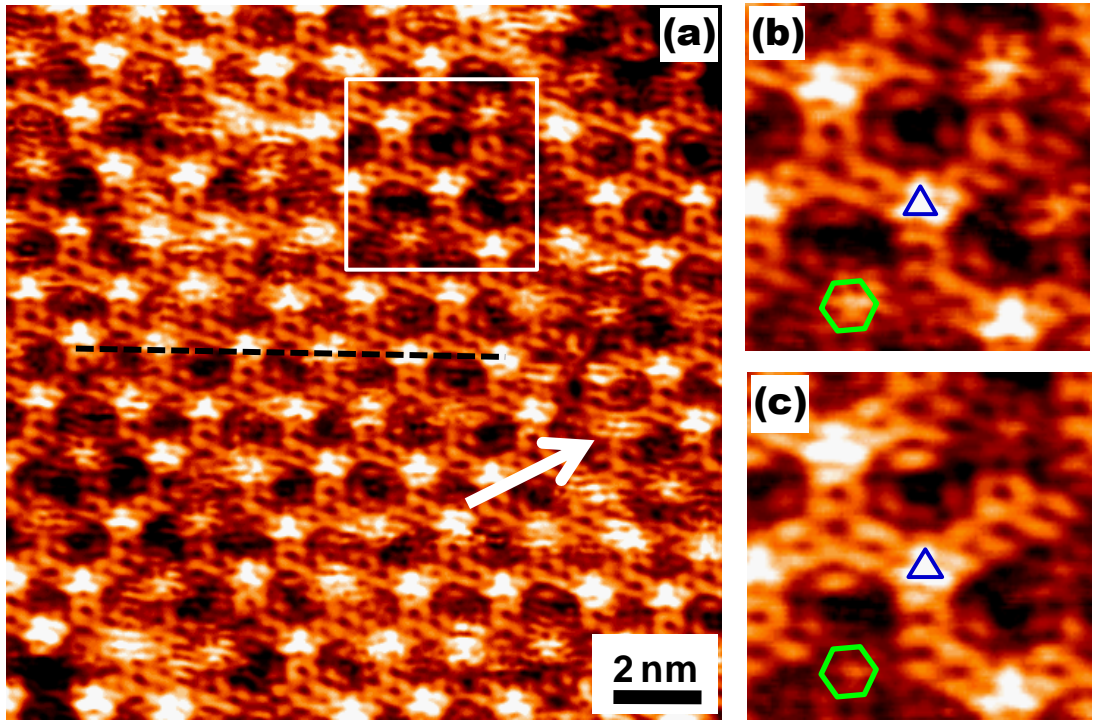

Qi et al. Fig. 1

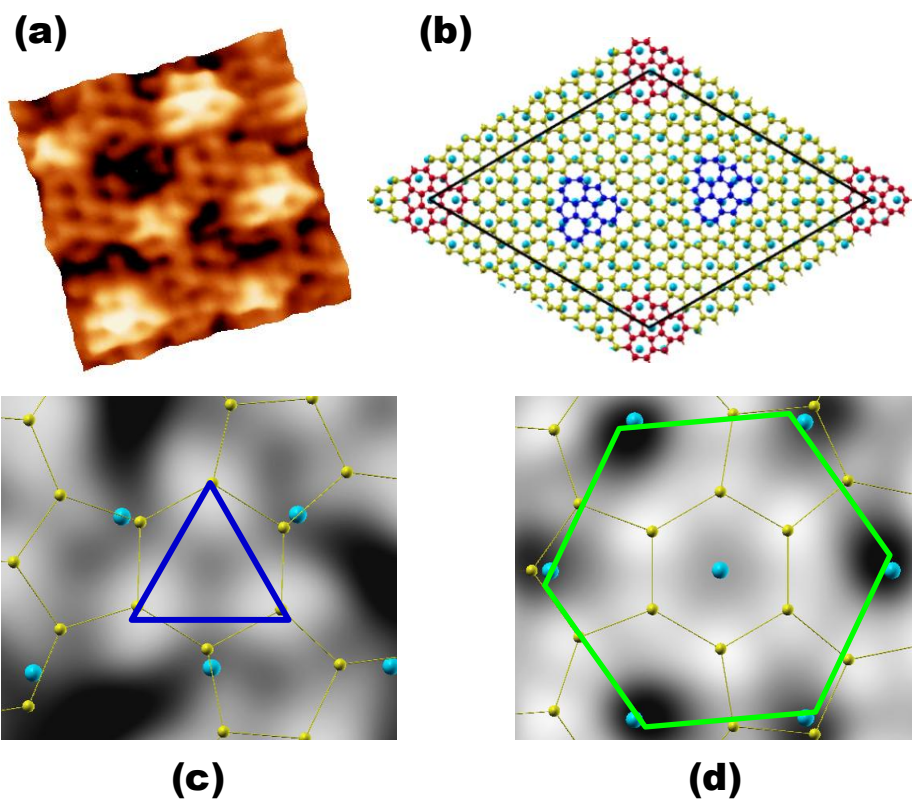

**(a)** **(b)**

**(c)** **(d)**

Qi et al. Fig. 2

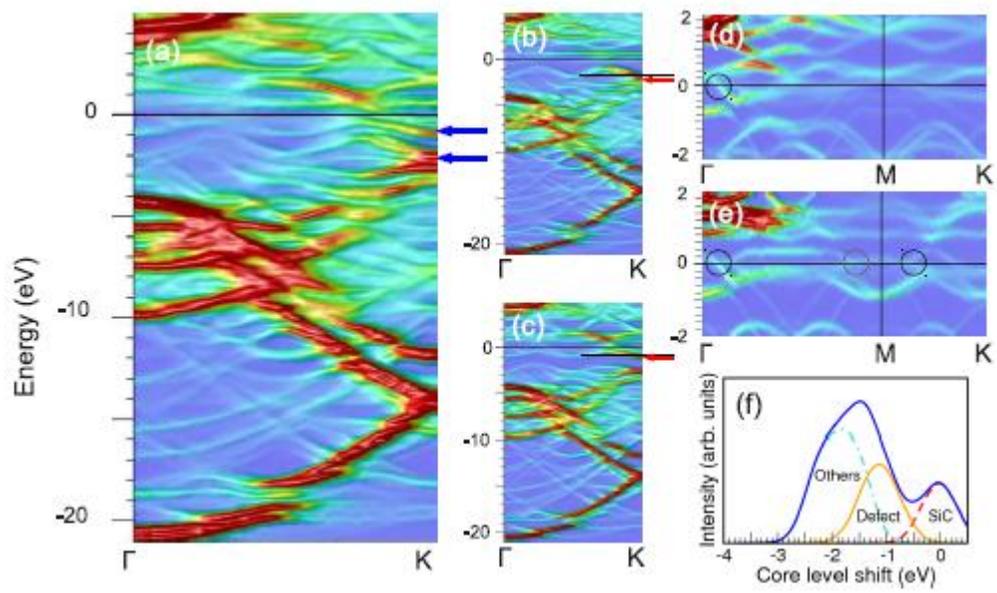

Qi et al. Fig. 3

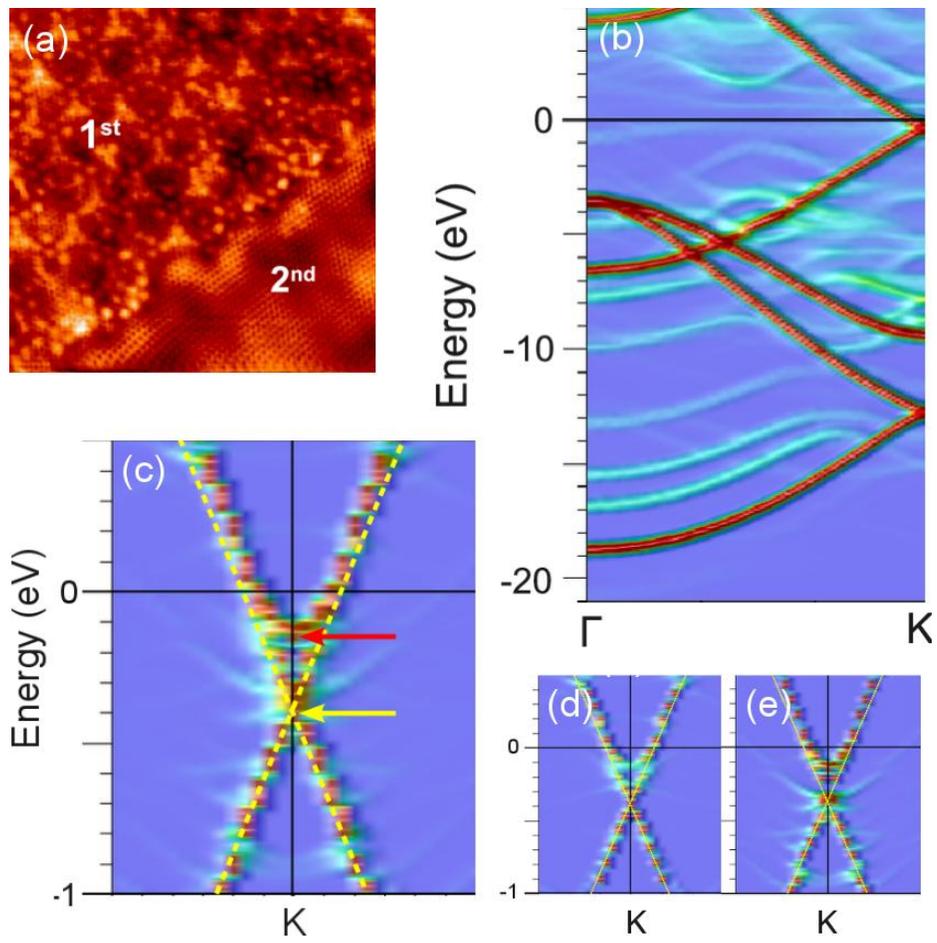

Qi et al. Fig. 4